\documentclass[aps,showpacs,prd,twocolumn,superscriptaddress,floatfix]{revtex4}
\usepackage{amsmath, amsfonts, hyperref}
\usepackage{color}
\usepackage{graphicx}
\newcommand{\beq}{\begin{equation}}
\newcommand{\eeq}{\end{equation}}
\newcommand{\beqn}{\begin{eqnarray}}
\newcommand{\eeqn}{\end{eqnarray}}

\def\vhat{\hat v}
\def\vhats{\hat v_*}
\def\Rcal{{\cal R}}
\newif\ifdraft
\drafttrue 

\begin{document}

\title
{The  Effect of Stars on the Dark Matter Spike Around a
Black Hole: \\ A Tale of Two Treatments}
\date{\today}
\author{Stuart~L. Shapiro}
\affiliation{Department of Physics, University of Illinois at Urbana-Champaign, Urbana, IL 61801, USA}
\altaffiliation{Also Department of Astronomy and NCSA, University of
  Illinois at Urbana-Champaign, Urbana, IL 61801, USA}
\author{Douglas C. Heggie}
\affiliation{School of Mathematics and Maxwell Institute for Mathematical Sciences, University of Edingburgh, Kings Buildings, Edinburg EH9 3FD, UK}

\begin{abstract}
	We revisit the role that gravitational scattering
	off stars plays in establishing the steady-state 
	distribution of collisionless dark matter (DM) 
	around a massive black hole (BH).
	This is a physically interesting problem that 
	has potentially observable signatures, such as
	$\gamma-$rays from DM annihilation in a density spike. 
	The system serves as a laboratory for comparing two different
	dynamical approaches, both of which have been widely used: 
	a Fokker-Planck treatment and a two-component conduction fluid 
	treatment. In our Fokker-Planck analysis we extend a 
	previous analytic model to 
	account for a nonzero flux of DM particles into the BH,
	as well as a cut-off in the distribution function near the BH
	due to relativistic effects or, further out, possible
	DM annihilation. In our two-fluid analysis,
	following an approximate analytic treatment, we recast the
	equations as a ``heated Bondi accretion" problem and solve
	the equations numerically without approximation. While both 
	the Fokker-Planck and two-fluid methods yield basically the same 
	DM density and velocity dispersion profiles away from the 
	boundaries in the spike interior, there are other differences, 
	especially the determination of the DM accretion rate. 
	We discuss limitations of the two treatments,
	including the assumption of an isotropic velocity dispersion.

\end{abstract}

\maketitle

\section{Introduction} 

A supermassive black hole (SMBH) will steepen the density profile of
dark matter (DM) within the hole's sphere of influence, i.e., within
radius $r_h = GM_{bh}/v^2_0$.  Here, $M_{bh}$ is the mass of the hole and 
$v_0$ is the velocity dispersion in the galaxy core.  The
density profile of this DM spike depends both on the
properties of DM and the formation history of the SMBH. If the DM is
collisionless with a cuspy, spherical, inner halo density that follows a
generalized Navarro-Frenk-White (NFW~\cite{NavFW97}) profile then the
density in the absence of the hole will obey a power-law profile,
$\rho(r) \sim r^{-\gamma_c}$.  Simulations with DM alone yield typical
powers of $0.9 \lesssim \gamma_c \lesssim
1.2$~\cite{DieKMZMPS08,NavLSWVWJ10}, but if baryons undergo dissipative
collapse into { a baryonic} disk 
they can induce the adiabatic contraction of
the central DM halo into a steeper power
law~\cite{BluFFP86,GneKKN04,GusFS06}, with values as high as $\gamma_c
\sim 1.6$ allowed for our Galaxy~\cite{PatIB15}.

If the SMBH grows adiabatically from a smaller seed~~\cite{Pee72a} the
SMBH then alters the profile inside $r_h$, forming a DM spike within
which $\rho(r) \sim r^{-\gamma_{\rm sp}}$, where $\gamma_{\rm sp}=
(9-2\gamma_c)/(4-\gamma_c)$~\cite{GonS99}.  For $0 < \gamma_c \leq 2$
the power-law $\gamma_{\rm sp}$ varies at most between 2.25 and 2.50
for this case. However, gravitational scattering off of a dense
stellar component inside $r_h$ could heat the DM, softening the spike
profile and ultimately driving it to a final equilibrium value of
$\gamma_{\rm sp} = 1.5$~\cite{Mer04,GneP04,MerHB07}, or even to
disruption~\cite{WanBVW15}.  Other spikes,
characterized by other power laws, are obtained for alternative
formation histories for the BH within its host halo, such as the
sudden formation of a SMBH through direct collapse of gas inside DM 
halos~\cite{BegVR06}, mergers or gradual growth from an
inspiraling off-center seed \cite{UllZK01}, or in the presence of DM
self-interactions~\cite{ForB08,ShaP14,FenYZ21}. It is also possible
that baryon clumps can erase the DM density cusp via dynamical 
friction~\cite{Elz01,Rom08}.

DM annihilations in the innermost region of the spike, if they
occur, weaken the density profile there.
The density continues to rise with decreasing 
distance $r$ from the BH, as it forms a ``weak cusp''~\cite{Vas7,ShaS16} 
rather than a plateau~\cite{GonS99}. Within
the weak cusp the density increases as $r^{-1/2}$ for $s$-wave DM
annihilation and somewhat more slowly for $p$-wave annihilation.

Due to their very high DM densities, BH-induced density
spikes can appear as very bright gamma-ray point sources in models of
annihilating DM~\cite{GonS99,Mer04,GneP04,GonPQ14,FieSS14,BelS14,LacBS15,SheSF15}. 
Many of these models are now becoming detectable with
current and near-future high-energy gamma ray experiments,
and indeed the excess of $\sim 1-5$ GeV gamma rays from the inner 
few degrees of the Galactic Center (GC) observed by {\it Fermi} 
may prove to be a first signal of annihilating 
DM~\cite{DayFHLPRS16,CalCW15,FermiGCE16},
although tension with limits from dwarf galaxies~\cite{FermiDwarf15}
and the statistical properties of the photons in the GC excess
~\cite{LeeLSSX16,BarKW16} may indicate a more conventional 
astrophysical explanation for
the GC excess, such as a new population of pulsars (see, e.g.~,
\cite{AbaCH14, BraK15, OleKKD16}).

Here we revisit the issue of Newtonian 
gravitational scattering of collisionless DM
off a stellar component inside $r_h$ in the presence of a massive, 
central BH. Our motivation is multipurpose: (1) to
obtain the steady-state profile of DM in the cusp to which the time-dependent,
numerical integration of the Fokker-Planck equation in \cite{Mer04} asymptotes 
at late times; (2) to generalize the zero-flux, steady-state solution of 
the Fokker-Planck equation in \cite{GneP04} to allow for a net flux 
of DM onto the BH; and, especially, 
(3) to use this problem as one of the simplest laboratories 
that can be exploited to compare a Fokker-Planck approach
to a two-fluid conduction approach for treating the dynamical 
behavior of a two-component cluster of collisionless gases 
interacting by gravitational scattering 
alone~(see, e.g. \cite{BetI85,HegA92,SpuT95} and references therein). 
Our Fokker-Planck
treatment is entirely analytic. Our two-fluid conduction treatment
is first performed analytically to gain insight, 
after we adopt some reasonable approximations. 
Then, once we recast the DM fluid 
equations in the form of a ``heated Bondi accretion" problem, we 
solve them numerically without approximation.

The plan of the paper is as follows. In Section II we present
our Fokker-Planck treatment and in Section III our two-component
fluid treatment. In Section IV we discuss some of the 
implications of our dual analyses. 
We adopt gravitational units and set $G=1=c$ throughout.

\section{Fokker-Planck Treatment}
\label{FPT}

\subsection{Phase-Space Distribution Function}

We begin by following ~\cite{Mer04,GneP04} and adopting 
a Fokker-Planck approach to addressing the problem. 
We regard a Fokker-Planck treatment 
as the more fundamental approach (compared with a fluid approach) to analyzing 
Newtonian N-body systems that evolve by 
undergoing cumulative, small-angle gravitational (Coulomb) 
scatterings on two-body relaxation timescales.
Here we have a two-component system consisting of DM 
particles that scatter off stars to establish a (quasi)stationary 
DM distribution in the presence of a massive, central black hole (BH) 
of mass $M_{bh}$ that dominates the potential in the spike.
The Fokker-Planck equation can be employed to evolve the 
phase-space distribution 
function $f(E,t)$ of DM particles bound to the BH in the spike, where
$E = M_{bh}/r-v^2/2 > 0$ is the DM binding energy per unit mass. Here
$r$ is the radius from the BH and $v$ is the speed of a particle; 
the velocity dispersions are
assumed isotropic for both DM particles and stars.
A power-law distribution function for the DM 
satisfying $f(E) \sim E^p$ gives rise to a power-law DM
density, $\rho \sim r^{-3/2-p}$. 

The Fokker-Planck equation for the evolution of the 
distribution function $f$ of DM particles of mass $m_{\chi}$ in the presence
of stars of mass $m_*$ can be written in the form ~\cite{Mer04,GneP04} (see
also ~\cite{Spi87}, Eq. 2-86, with a slight change of notation)
\begin{align} 
\label{FPE}
-\frac{\partial q(E)}{\partial E}\frac{\partial f}{\partial t}
	&= A \frac{\partial}{\partial E} \left[\frac{m_{\chi}}{m_*} f
\int_E^\infty f_* \frac{\partial q_*}{\partial E_*}dE_*\right.  \\   
&+\left. \frac{\partial f}{\partial E}  \Bigg\{\int_E^\infty f_* q_* dE_*   
+ q \int_{-\infty}^E f_* dE_* \Bigg\} \right], \nonumber
\end{align}
where  
$q(E) = (2^{-1/2}/3)\pi M_{bh}^3E^{-3/2}, 
A \equiv 16 \pi^2 m_*^2 \ln \Lambda$ and $\ln \Lambda = \ln (M_{bh}/m_*)$.
Here $f_*$ is the distribution function of the stars, which we take
to be a fixed power-law in the cusp,
\begin{equation}
  f_* = K E^s, E>0
  \label{fstar}
\end{equation}
for this exercise. The constant $K$ determines the magnitude of the 
stellar density at a fiducial point in the 
cusp (see below) and the power-law with 
$-1 < s < 1/2$ determines the density 
profile there, $\rho_* \sim r^{-\beta}$, where $\beta = s+3/2$. 
For this analysis we set the stellar density to be zero for
unbound stars that orbit outside the cusp: $f_* = 0,$ $E < 0$.
The equilibrium distribution function we might 
expect for the bound stars is  
$f_* \propto E^{1/4}$, i.e. $s=1/4$, corresponding to 
$\rho_* \propto r^{-7/4}$, which is the Bahcall-Wolfe (BW)~\cite{BahW76} 
steady-state solution for a one-component, isotropic system of 
stars deep inside the cusp around a massive BH. 
However we shall leave 
$s$ and $\beta$ unspecified in what follows. In principle, it is
determined by solving the Fokker-Planck equation for the stars in 
conjunction with Eq.~\ref{FPE} for the DM.

For DM particles the first term in square brackets in Eq.~(\ref{FPE})
is negligible since $m_\chi/m_* \ll 1$.
Also we can recast Eq.~(\ref{FPE}) as a continuity equation in $E$-space, as follows. 
Consider the DM particle number density per unit energy,
$N(E,t) =  4 \pi^2 p(E) f(E,t)$, where 
\begin{eqnarray}
p(E) 
	&\equiv& 4 \int_0^{r_{{\rm max}(E)}} v r^2 dr 
= -\partial q(E)/\partial E \\ 
	&=& 2^{-3/2} \pi M_{bh}^3 E^{-5/2}, \nonumber
\end{eqnarray}
where $r_{max}(E) = M_{bh}/E$ is the maximum radius reached by a particle
orbiting with energy $E$.
Then Eq.~(\ref{FPE}) becomes
\begin{equation} 
\label{FP2}
	4 \pi^2 p(E)\frac{\partial f}{\partial t}
	= \frac{N(E,t)}{\partial t}
	= -\frac{\partial {\cal F}(E,t)}{\partial E}.
\end{equation}
Here the particle flux in E-space, ${\cal F}(E,t)$, is given by 
${\cal F}(E,t) \equiv -4 \pi^2 A \frac{\partial f}{\partial E} \{ \ \}$, 
where the terms inside the curly brackets $\{ \ \}$ are the terms in 
curly brackets on the right-hand side of Eq.~(\ref{FPE}).

An equilibrium solution 
satisfying $\partial f/\partial t =0$ with {\it no
energy flux} then requires $\partial f/\partial E = 0$, or $p=0$.
The resulting density profile is then $\rho \propto r^{-3/2}$.
This simple argument for the DM spike was first presented in 
~\cite{GneP04}.  What is particularly interesting, as the above 
derivation demonstrates, is that this steady-state DM density profile 
arises {\it independently} of the assumed background 
stellar distribution function, $f_*(E)$, in the zero-flux case. 
This same DM equilibrium solution was also achieved at late times, 
away from the cusp boundaries, in the time-dependent, 
numerical integration reported in ~\cite{Mer04}.  There 
Eq.~(\ref{FPE}) was evolved, starting from an 
adiabatic DM spike with $\rho \propto r^{-7/3}$  
in a fixed background stellar density cusp,
after adding an additional flux term to mimic the expected additional 
capture of DM particles scattered into the black hole loss-cone, 
were the restriction to isotropy relaxed ~\cite{FraR76,LigS77} (see
discussion in Section II.D below).

We now generalize the derivation in \cite{GneP04} by allowing for
a {\it nonzero} energy flux, since DM particles may be captured by the 
BH even for an isotropic distribution.
Evaluating the two integrals in the curly brackets in Eq.~(\ref{FPE}) and 
seeking a steady-state solution again by setting 
$\partial f/ \partial t = 0$ implies 
$\partial {\cal F}(E)/\partial E =0$, or
\begin{equation}
\label {eq0}	
\frac{d}{dE}\left[ E^{s-1/2} \frac{df}{dE} \right] = 0,	
\end{equation}
whose solution is
\begin{equation} 
\label{eq}
f(E) = \frac{F}{(3/2-s)}E^{3/2-s}+C,	
\end{equation}	
where $F$ and $C$ are constants. Substituting Eq.~(\ref{eq}) into the
definition of the particle flux ${\cal F}(E)$ shows that $F$ is related
to ${\cal F}(E)$ according to
\begin{equation}
\label{F}
{\cal F}(E) = \frac{6 \pi^2 A {\tilde q} K}{(s+1)(s-1/2)} F =
	\ \ {\rm constant},
\end{equation}
where we introduced another constant 
${\tilde q} \equiv q(E)E^{3/2}$. Eq.~(\ref{F}) shows that 
${\cal F}(E)$ is 
constant in both $t$ and $E$. The two constants $F$ (or ${\cal F}(E))$
and $C$ are determined by two boundary conditions that we can impose
on Eq.~\ref{eq}:
\begin{eqnarray}
  \label{first-bc}
&{\rm b.c.}& \ (i): \ \ \ f = 0, \ \ E > E_{cut} \equiv M_{bh}/r_{cut}, \\    
&{\rm b.c.}& \ (ii): \ \ \rho(r) = \rho_h, \ \ r = r_h = M_{bh}/v_0^2, \nonumber
\end{eqnarray}
The first boundary condition cuts off the DM distribution function 
for high  energies characterizing DM orbits that would otherwise 
reside entirely very near the BH.
For example, any particle that penetrates the marginally bound
radius, where  $r_{mb} = 4 M_{bh} \ll r_h$ for a Schwarzschild BH, 
must plunge directly into the BH (see, e.g. the discussion in 
\cite{ShaP14} and references therein). In this case we should set 
$r_{cut}=r_{mb}$. Of course, relativistic effects
would modify our Newtonian treatment in this region, but including them 
is beyond the scope of this analysis and does not affect our main results
at larger radii. Alternatively, if our DM particles were to undergo
annihilation reactions within a larger domain 
$r_{mb} < r \leq r_{ann}$, 
then we must set $r_{cut}=r_{ann} \ll r_h$ \cite{Vas7,ShaS16,FieSS14}. 

The second boundary condition sets the DM density to a fiducial value
$\rho_h$ at the outer boundary of the spike, where the density can be
inferred by, e.g., extrapolating from solar neighborhood 
estimates in the case of the Galaxy (see Section IIB below). 
Inserting b.c.~(i) into
Eq.~(\ref{eq}) allows us to relate $F$ and $C$,
\begin{equation}
\label{FC}
	F=-\frac{(3/2-s)}{E_{cut}^{3/2-s}} C.	
\end{equation}
Substituting 
Eq.~(\ref{eq}) into the relation for the DM density, 
\begin{eqnarray}
\label{rhoDM0}
\rho(r) &=& m_{\chi} \int_0^{E_{\rm max}} 4 \pi v^2 f(E) dv \\
	&=& 4\pi m_{\chi} \int_0^{M_{bh}/r} [2(M_{bh}/r -E)]^{1/2} 
	f(E)dE.  \nonumber 
\end{eqnarray}
yields $\rho(r)$ vs. $r$
in terms of $F$ and $C$. Then employing b.c.~(ii) and 
evaluating $\rho$ at $r=r_h$ yields
a second relation between $F$ and $C$ in terms of $\rho_h$.
Using both of these relations for $F$ and $C$ then allows us to evaluate 
Eq.~(\ref{eq}) for $f(E)$ in terms of $\rho_h$ and $E_{cut}$:
\begin{equation}
	\label{f(E)}
f(E) = 
	\frac{(\rho_h/m_{\chi})(M_{bh}/r_h)^{-3/2}}{\frac{2^{7/2} \pi}{3}  
	(1-\frac{3}{2}(r_{cut}/r_h)^{3/2-s}I)}
	{\left[ 1-\left( \frac{E}{E_{cut}} \right)^{3/2-s} \right]},
\end{equation}
where $I=B(5/2-s,3/2)$ and $B(x,y)$ is the standard beta function,
i.e. $I=\int_0^1 dx (1-x)^{1/2} x ^{3/2-s}$.

\subsection{Density}

Inserting Eq.~(\ref{f(E)}) into (\ref{rhoDM0}) yields the DM density
profile,

\begin{align}
\label{densfok}
\frac{\rho(r)}{\rho_h} =& 
\frac {1 - \frac{3}{2}(\frac{r_{cut}}{r})^{3/2-s}I} 
{1 - \frac{3}{2}(\frac{r_{cut}}{r_h})^{3/2-s}I}
	\left( \frac{r_h}{r} \right)^{3/2}, \ r \geq  r_{cut} \\
          =&
\frac {1-(1-\frac{r}{r_{cut}})^{3/2} - 
	\frac{3}{2}(\frac{r_{cut}}{r})^{3/2-s}{\cal I}}
{1 - \frac{3}{2}(\frac{r_{cut}}{r_h})^{3/2-s}I}
	\left( \frac{r_h}{r} \right)^{3/2}, 
	\ r < r_{cut} \nonumber 
\end{align}
where ${\cal I} = B(r/r_{cut};5/2-s,3/2)$, 
and where $B(x;a,b)$ is the standard incomplete beta function; 
more transparently, 
${\cal I} = \int_0^{r/r_{cut}} dx (1-x)^{1/2} x^{3/2-s}$.

\begin{figure}
\includegraphics[width=7cm]{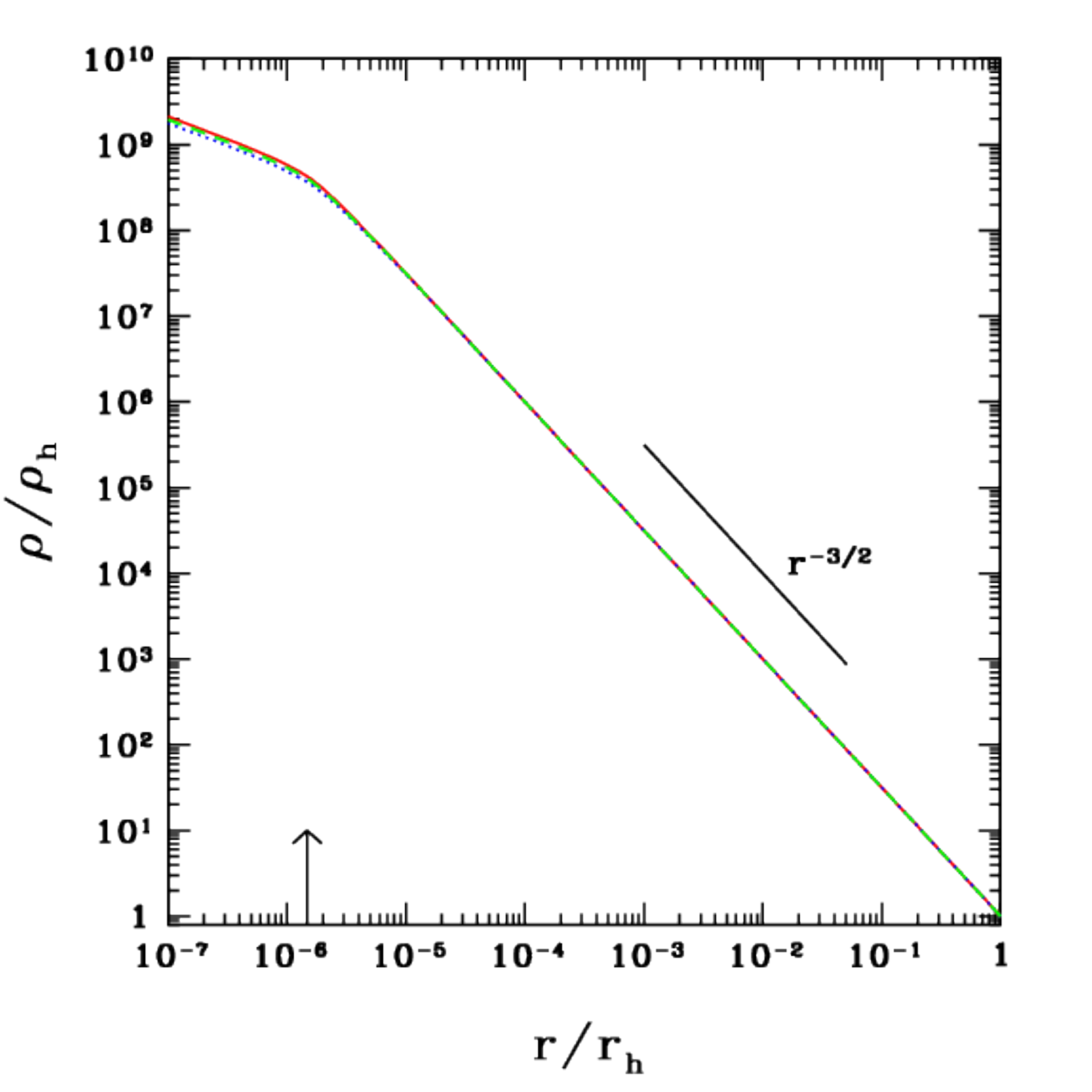}
\includegraphics[width=7cm]{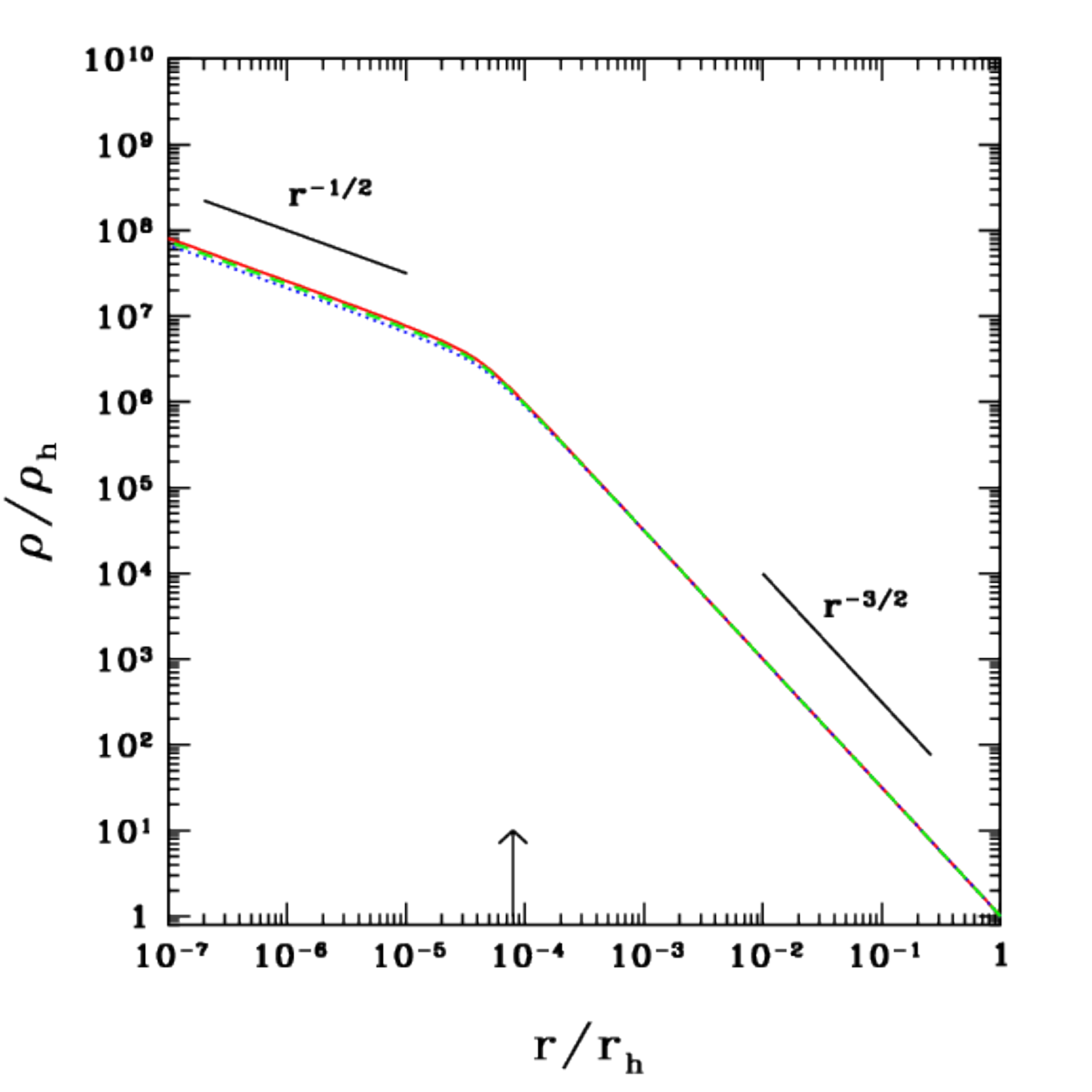}
\caption{Fokker-Planck solution for the DM steady-state density 
profile $\rho(r)$ in the spike around a massive black hole, 
allowing for background stars. The DM distribution function cuts off 
at $r_{cut} = r_{mb}$ (upper figure) 
and $r_{ann}$ (lower figure); vertical arrows show the location
of $r_{cut}$. Three stellar density profiles
$\rho_* \sim r^{-\beta}$ are chosen for each figure: 
$\beta=$1 (solid, red); 7/4 (dotted, blue); 1.4 (dashed, green).
The three curves are nearly indistinguishable in the plot.
The densities and radii are normalized to their values near the
spike outer boundary at $r_h$. Parameters are chosen that characterize
a spike around Sgr A* in the Galactic center (see text).
}
\label{fig:densfok}
\end{figure}

We evaluate the density profile given by Eq.~(\ref{densfok}) and plot
the results in Fig.~\ref{fig:densfok}. We consider two cases for
$r_{cut}$: one in which $r_{cut} = r_{mb}$ (upper plot) and the other
in which $r_{cut} = r_{ann}$ (lower plot). For each case we treat
three possibilities for the power-law profile of the background 
stars: $\beta = 1$ (NFW); $\beta = 7/4$ (BW) and $\beta = 1.4$
(Galactic center fit~\cite{Gen03,GneP04}). It is clear from the
form of the equation that for $r \gg r_{cut}$ the equilibrium 
DM density profile varies as $r^{-3/2}$, as in the zero-flux case and, 
as in that case, it does not depend at all on the stellar density. Moreover, 
for $r \lesssim r_{cut}$, the expression for the DM density 
only depends on the stellar phase-space distribution function power-law $s$
(or corresponding mass density power-law $\beta = s+3/2$)
and not its magnitude, and from the figure we see that even the
power-law dependence is barely noticeable.

In evaluating $r_{cut}$ in Fig.~\ref{fig:densfok} we
adopt parameters appropriate for a spike around Sgr A* in the Galactic 
center. Here $M_{bh} = 4 \times 10^6~M_{\odot}$~\cite{GenEG10,GheSWLD08},
giving $r_{mb} = 7.7 \times 10^{-7}$~pc. To estimate $r_{ann}$ we
follow \cite{FieSS14}, who adopt a self-conjugate
DM particle with mass $m_{\chi} = 35.25$~Gev annihilating to $b \bar b$
with a cross section $\langle \sigma v \rangle = 1.7 \times 
10^{-26}~{\rm cm^{3} s^{-1}}$
(typical WIMP values;~\cite{DayFHLPRS16}), and find
that the annihilation region sets in at $\rho_{ann} =
1.7 \times 10^8 ~M_{\odot} {\rm pc^{-3}} = 6.6 \times 10^9 
{\rm ~GeV cm^{-3}}$.
Taking the DM density in the solar neighborhood to be $\rho_D = 
0.008 ~M_{\odot} {\rm pc^{-3}} = 0.3~{\rm GeV cm^{-3}}$~\cite{BovT12}, and
$\rho_h = \rho_D (D/r_h)^{\gamma_c}, \gamma_c = 1$~(NFW), where 
$D=8.5$~kpc is the sun's distance to the Galactic center, 
we then find that $r_{ann} = { 4.4} \times 10^{-5}$~pc. 
Here we took $v_0 = 182~{\rm km s^{-1}}$ ($\sqrt 3$ times the 
line-of-sight velocity dispersion 
of $105~{\rm km s^{-1}}$~\cite{Guletal09}) 
to get $r_h = 0.52$~pc and
used Eq.~(\ref{densfok}) for the density inside $r_h$.

We see from Fig~\ref{fig:densfok} that the DM density departs 
significantly from $r^{-3/2}$ for $r \lesssim r_{cut}$.
This is a result of b.c.~(i) and is most evident for $r_{cut} = r_{ann}$,
where the density is seen to 
vary as $\rho \sim r^{-1/2}$ for $r \ll r_{cut}$. As discussed in
\cite{Vas7,ShaS16}, where this scaling was found previously, 
the particles occupying this region have energies much smaller than the 
potential there, and so they orbit with increasing eccentricity 
and apocenters as $r$ decreases below $r_{cut}$,
penetrating { well} within $r_{cut}$ only near pericenter.
Particles whose
orbits would reside entirely within $r_{cut}$ due to their large binding 
energy $E>E_{cut}$
are never present, as they would be destroyed by rapid capture
by the BH ($r_{cut} = r_{mb}$) or
annihilation ($r_{cut} = r_{ann}$), and this causes the 
reduction in the steepness of the density spike within $r_{cut}$.

\subsection{Velocity Dispersion}

The DM velocity dispersion may be computed from
\begin{equation}
	\label{velfok0}	
	v^2(r) = 
	 \frac{4 \pi m_{\chi}}{\rho(r)} 
	 \int_0^{M_{bh}/r} [2(M_{bh}/r -E)]^{3/2} f(E)dE.
\end{equation}
Inserting Eq.~(\ref{f(E)}) into (\ref{velfok0}) yields
\begin{align}
	\label{velfok}
	v^2(r) =& ~Q(r) 
\frac{1 - \frac{5}{2}(\frac{r_{cut}}{r})^{3/2-s}{\hat I}}
     {1 - \frac{3}{2}(\frac{r_{cut}}{r_h})^{3/2-s}I}, \ r \geq r_{cut} \\
	       =& ~Q(r)
	\frac{1 - (1-\frac{r}{r_{cut}})^{5/2} -
	\frac{5}{2}(\frac{r_{cut}}{r})^{3/2-s}{\hat {\cal I}}}
     {1 - \frac{3}{2}(\frac{r_{cut}}{r_h})^{3/2-s}I}, \ r < r_{cut} \nonumber
\end{align}
where
\begin{equation}
	\label{Q}
	Q(r)=\frac{6}{5} \left( \frac{M_{bh}}{r} \right) 
	\frac{\rho_h r_h^{3/2}}{\rho(r)r^{3/2}}.
\end{equation}
Appearing in the above equations are the two quantities 
${\hat I} = \int_0^1 dx (1-x)^{3/2} x^{3/2-s} = B(5/2-s,5/2)$ and also 
${\hat {\cal I}} = \int_0^{r/r_{cut}} dx (1-x)^{3/2} x^{3/2-s} = 
B(r/r_{cut};5/2 -s,5/2)$.

\begin{figure}
\includegraphics[width=6.8cm]{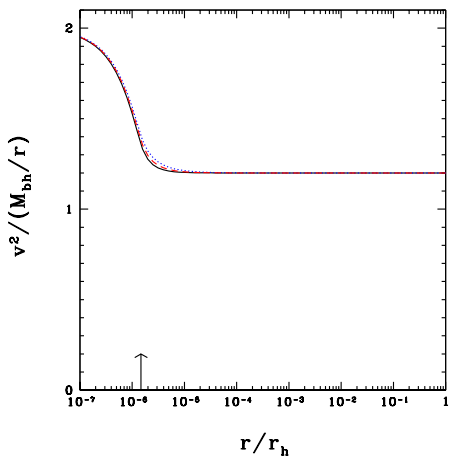}
\includegraphics[width=6.8cm]{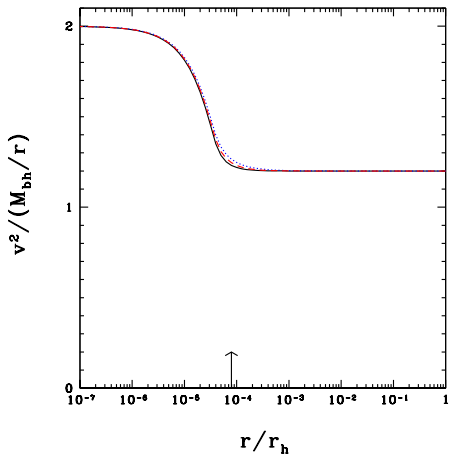}
\caption{Fokker-Planck solution for the DM steady-state velocity
dispersion profile in the spike around a massive black hole, 
allowing for background stars. Results are plotted for the cases
shown in Fig~\ref{fig:densfok} and the labelling is the
same as in that figure. The velocity dispersion is normalized
to the square of the local circular velocity $M_{bh}/r$.
}
\label{fig:vel2fok}
\end{figure}

We evaluate the velocity profile given by Eq.~(\ref{velfok}) for 
the cases shown in Fig.~\ref{fig:densfok} and plot the
results in Fig.~\ref{fig:vel2fok}. Once again the profiles do not
depend at all on the magnitude of the background stellar density 
and only insignificantly on the profile power-law $s$. 
As is seen most clearly in the
cases for which $r_{cut} = r_{ann}$, the DM velocity 
dispersion profile has two distinct
regimes. For $r \gg r_{cut}$ the profile is given by
$v^2/(M_{bh}/r) = 3/(p+5/2) = 6/5$, as expected for a DM distribution function
of the form $f(E) \sim E^p$ where $p=0$, or $\rho \sim r^{-3/2}$.
For $r \ll r_{cut}$ the profile
asymptotes to $v^2/(M_{bh}/r) = 2$, corresponding to $p=-1$, 
or $\rho \sim r^{-1/2}$; this result follows from the fact that 
this region is filled by $E \approx 0$ particles
in highly eccentric orbits near pericenter.

\subsection {Flux}

To evaluate Eq.~(\ref{F}) for the constant, nonzero DM flux $\cal F$ 
we must first determine $K$, defined in Eq.~(\ref{fstar}). This quantity serves to normalize the 
stellar distribution function $f_*$ to yield a specifed stellar 
density $\rho_{*h}$ at a fiducial radius, $r_h$. 
Employing an expression 
identical to Eq.~(\ref{rhoDM0}), but for stars
rather than DM particles, yields
\begin{equation}
\label{K}	
K=\frac{\rho_{*h}}{4 \pi 2^{1/2}  m_* (M_{bh}/r_h)^{s+3/2} {\tilde I}}
\end{equation}
where $\rho_{*h}$ is the stellar density at the spike boundary 
at $r=r_h$ and 
${\tilde I} = \int_0^1 dx (1-x)^{1/2} x^s = B(1+s,3/2)$.
Relating $F$ to $\rho_h$ and $E_{cut}$ as described below Eq.~(\ref{rhoDM0})
and inserting the result together with Eq.~(\ref{K}) into Eq.~(\ref{F}) 
yields the DM mass flux,
\begin{equation}
	\label{Fokflux}
	{\dot M} =-m_{\chi}{\cal F} = C_{\cal F} \frac {
		\ln{\Lambda} m_* {\rho_{*h}}{\rho_h}r_h^3} {(M_{bh}/r_h)^{3/2}}, 
\end{equation}

where
\begin{equation}
C_{\cal F} = \frac{(3/2-s)3\pi^3}{(s+1)(s-1/2)2^{3/2}{\tilde I}}
	\frac{(r_{cut}/r_h)^{3/2-s}}{[1-3/2(r_{cut}/r_hr)^{3/2-s}I]},
\end{equation} 
and where ${\tilde I}=\int_0^1 dx (1-x)^{1/2}x^s = B(1+s,3/2)$.
Using the local heating time for DM particles due to gravitational 
encounters with stars \cite{Mer04} ($\sim$ stellar relaxation time for 
distant, two-body encounters, assuming comparable stellar and DM 
velocity dispersions ), 
\begin{equation}
	\label{tr}
	t_r = \frac{0.0814 v^3}{m_* {\rho_*} \ln {\Lambda}}, 
\end{equation}	
allows us to recast Eq.~(\ref{Fokflux}) as 
\begin{equation}
	\label{fluxapprox}
	{\dot M} \sim \frac{M_{DM}}{t_{rh}} 
		    \left(\frac{r_{cut}}{r_h}\right)^{3/2-s} 
		    \ll \frac{M_{DM}}{t_{rh}},
\end{equation}
where $M_{DM} \sim 4 \pi r_h^3 \rho_h/3$ is the total DM mass
inside the spike and $t_{rh}$ is the relaxation time at $r=r_h$.

The mass flux given in Eq.~(\ref{fluxapprox}) is reminescent of the
BW solution for the steady-state mass flux for stars onto a
central black hole. BW also assumed that the distribution function
was of the form $f(E,t)$, representing an isotropic system. The flux
at late times was found to asymptote to the steady-state 
value~\cite{BWflux} 
\begin{equation}
	\label{fluxBW}
	{\dot M^*_{BW}} \sim \frac{M_{*}}{t_{rh}} 
                    \left(\frac{r_{cut}}{r_h}\right)^{3/2-2s} 
                    \ll \frac{M_{*}}{t_{rh}},
\end{equation}
with $M_* \sim 4 \pi r_h^3 \rho_{*h}/3$ and $s=1/4$. 
The difference between the exponent $s$ in 
Eq.~(\ref{fluxapprox}) for the DM flux and $2s$ in Eq.~(\ref{fluxBW}) for 
the stellar flux is due to the fact that the flux of DM is 
driven by interactions with 
background stars while the flux of stars is driven by 
self-interactions with other stars.

A key point to appreciate
is that Eqs.~(\ref{fluxapprox}) and ~(\ref{fluxBW})
are both wrong! When proper allowance is made for an {\it anisotropic}
velocity
dispersion described by a distribution function of the form $f(E,J)$, 
where $J$ is the angular momentum per unit mass, 
it turns out that the correct flux is much larger,
\begin{equation}
	\label{flux2d}
	{\dot M_{lc}} \sim \frac{M_h}{t_{rh}}, \ \ \ f=f(E,J) \neq f(E), 
\end{equation}
where $M_h = M_{DM}$,
or $M_h=M_*$, depending on the component. The reason is that the
bulk of the flux originates from  high-eccentricity orbits in the outer
cusp that are scattered into the black hole {\it loss-cone}
and captured in an orbital period. While the loss-cone only breaks
isotropy logarithmically, it significantly increases the capture rate.
This result was first shown analytically in \cite{FraR76,LigS77} 
and confirmed in more detail
numerically in \cite{ShaM78,CohK78}. [For an early review and 
references, see ~\cite{Sha85}].  An extra 
sink term that roughly accounts for
the DM loss-cone capture rate was inserted 
in the Fokker-Planck equation for $f(E,t)$ in  ~\cite{Mer04},
similar to the ``patch" introduced in an earlier treatment of 
the equilibrium stellar distribution performed in ~\cite{LigS77} 
for $f(E)$ as a follow-up to 
the more general analysis of $f(E,J)$ in that paper. 
This sink term does not change the equilibrium density or 
velocity dispersion profile significantly. 
Generalizing the isotropic analysis presented here
by solving instead for an anisotropic DM distribution 
function of the form $f(E,J)$  
is possible, but not the purpose of this paper. The results
should confirm those anticipated above with regard to the role of
the loss-cone. Most importantly, 
while the flux would be
significantly increased by allowing for the associated anisotropy,
the modification of the density and velocity profiles would not
be significant, as the deviation in $f$ from isotropy would only 
consist of a slowly-varying logarithmic function of $J$ that reduces
the DM distribution as one approaches the 
loss cone at low-$J$~{\cite{FraR76,LigS77}}.

For the Galactic center we consider stars at $r_h$ with 
$\rho_{*h} = 1.2 \times 10^6~M_{\odot} {\rm pc}^{-3}$, $m_* = M_{\odot}$
and $\Lambda = 0.4N, N \approx 6 \times 10^6$~\cite{Mer04}. Together with the
adopted parameters for the DM listed above, we then have
$t_{rh} \approx 1.5 \times 10^9~$yrs and a DM mass inside $r_h$ of
$M_{DM} \approx 80~M_{\odot}$,
which gives an anticipated  DM accretion rate from Eq.~(\ref{flux2d}) of
$\dot M \sim 5 \times 10^{-8}~M_{\odot}{\rm yr}^{-1}$.

\section{Two-Component Fluid Treatment}

Adapting the two-component fluid formalism presented 
in ~\cite{BetI85,HegA92} to the problem at hand, the fluid 
equations for the DM particles analogous to Eq.~(\ref{FPE}) become 
\begin{equation}
\label{cont}
\frac{\partial \rho}{\partial t} + 	
\frac{1}{r^2}\frac{\partial (\rho u r^2)}{\partial r} = 0,  
\end{equation}
\begin{equation}
\label{mom}
\frac{\partial u}{\partial t} + u \frac{\partial u}{\partial r} = \\
-\frac{1}{\rho} \frac{\partial P}{\partial r} - \frac{M_{bh}}{r^2}, 
\end{equation}
\begin{eqnarray}
\label{energy}
	&&4 \pi r^2 \rho \vhat^2 \left(\frac{D}{Dt}\right) 
	\ln \left(\frac{\vhat^3}{\rho}\right) = 4 \pi r^2 \Gamma 
	\equiv \\ 	
&& 16 \pi r^2 (2 \pi)^{1/2} \ln \Lambda 
\left[\frac{\rho \rho_*}{(\vhat^2 + \vhats^2)^{3/2}}\right] 
	(m_* \vhats^2 -m_{\chi} \vhat^2). \nonumber
\end{eqnarray}

In the above equations, $u$ is the mean 
radial velocity and the pressure $P = \rho \vhat^2$, where $\vhat$ is the
one-dimensional (i.e., line-of-sight) velocity dispersion.
The dispersion is again assumed isotropic, whereby 
$\vhat^2 = v^2/3$, and similarly for the stars 
(i.e. $\vhats^2 = v_*^2/3$). The Lagrangian time derivative
$D/Dt$ my be expanded in the usual way according to 
$D/Dt = \partial/\partial t + u \partial/\partial r$.
The quantity $\Gamma$ appearing in Eq.~(\ref{energy}) gives the
DM heating rate per unit volume by gravitational scattering 
off stars~\cite{Spi87}. The other variables appearing above
have their same meanings as in Section II.
Once again we assume that the background stellar profile is fixed
and given by a power-law with $\rho_* \sim r^{-\beta}$. 

The equation of state is a $\gamma$-law with $\gamma = 5/3$, 
as it can be written in the form 
$P=(\gamma-1) \rho \epsilon$, where
$\epsilon = 3\vhat^2/2$ is the particle energy per unit mass.  
This identification is usually of no significance for 
applications involving nonrelativistic particles, such as stars in Newtonian 
stellar dynamics, but 
it will be useful below in drawing an analogy with the 
theory of Bondi accretion.

We note that $\Gamma$ has been derived assuming local Maxwellian
velocity distributions for both components. The basic 
functional dependence of this term on the local density and velocity
dispersion should be the same for other velocity distributions even though 
the numerical coeffients may change. This fact should be sufficient
to give the correct scaling of the $\rho$ and $v$ profiles with $r$
even for non-Maxwellian distributions, as found in previous
dynamical studies (see, e.g.,\cite{LigS77,You77}).

We are interested in solving the above fluid equations for
steady-state, hence we can drop all terms 
involving $\partial/\partial t$.
In addition, we can drop the term on the right-hand side of
Eq.~(\ref{energy}) involving $m_{\chi}$, 
as $m_{\chi} \vhat^2 \ll m_* \vhats^2$.
The resulting equations then become
\begin{equation}
\label{cont2}
	4 \pi r^2 \rho u = {\dot M} = {\rm constant},
\end{equation}
\begin{equation}
\label{mom2}
u \frac{d u}{d r} =
-\frac{1}{\rho} \frac{\partial P}{\partial r} - \frac{M_{bh}}{r^2},
\end{equation}
\begin{equation}
\label{energy2}
	-\frac{d \ln (\vhat^3/\rho)}{d \ln r}=
\frac{4 \pi \Gamma r^3}{{\dot M} \vhat^2} = 
	\frac{3}{2}\Rcal,
\end{equation}
where
\begin{eqnarray} 
	\label{RcalE}
\Rcal &\equiv& (\Gamma \frac{4}{3} \pi r^3)/
			 (\frac{1}{2}{\dot M} \vhat^2) \\
      &\approx& \frac{\rm heating \ rate \ inside \ r \ by \ stars}
{\rm  \ heat \ per \ unit \ time \ transported \ across \ r}. \nonumber
\end{eqnarray}
Note that in Eq.~(\ref{cont2}) and below we take $u$ to be 
the {\it magnitude} of the (inward) radial velocity.
In obtaining Eq.~(\ref{energy2}) we substituted Eq.~(\ref{cont2}) 
into Eq.~(\ref{energy}). If everywhere $\Rcal \ll$ 1 then the heating
of DM by gravitational scattering off stars is unimportant and the
DM gas is adiabatic (specific entropy $s \propto \ln (\vhat^3/\rho) =$ constant) and reduces to adiabatic 
Bondi flow~\cite{Bon52} for $\gamma = 5/3$.

We solve equations~(\ref{cont2}-\ref{energy2}) numerically without 
approximation in Section IIIB. In the next section we introduce a few
simplificatins that enable us to solve them analytically to gain
some preliminary insight.

\subsection{Approximate Analytic Solution}

We anticipate that the mean flow will be highly subsonic
($u \ll a$, where $a = \sqrt{\gamma P/\rho}$ is the DM sound speed), 
whereby we can eliminate the advective term on the 
left-hand side of the momentum equation~(\ref{mom2}). 
With this simplification (\ref{mom2}) reduces to the 
equation of hydrostatic equilibrium,  
\begin{equation}
\label{hydroeq}
	\frac{dP}{dr} = -\frac{M_{bh}}{r^2} \rho.
\end{equation}

Next, if we neglect heating ($\Rcal = 0$) and seek power-law solutions, 
Eqs.~(\ref{energy2}) and (\ref{hydroeq}) give
  \begin{equation}
    \rho = \rho_h(r_h/r)^{3/2}, {\vhat}^2 = {\vhat_h}^2(r_h/r).
    \label{exact1}
  \end{equation}
  Similarly, the assumed stellar density distribution gives the corresponding expressions
  \begin{equation}
    \rho_* = \rho_{*h}(r_h/r)^{\beta}, v_*^2 =     v_{*h}^2(r_h/r).
    \label{exact2}
  \end{equation}

  Now we treat the heating as a small perturbation on these exact expressions.  For this purpose we can substitute Eqs.~(\ref{exact1}) and (\ref{exact2}) into the perturbation term on the right of Eq.~(\ref{energy2}), finding
  \begin{equation}
\Rcal = \Rcal_h (r/r_h)^{3-\beta},    \label{Rcal}
  \end{equation}
 where ${\Rcal}_h = {\Rcal}(r_h)$.
  Now the perturbations to the results in Eq.~(\ref{exact1}) can be found.  Details are given in Appendix \ref{A1}, and lead to the result that
\begin{align}
 \label{rhoana}
 \frac{\rho}{\rho_h} &= \left(\frac{r_h}{r}\right)^{3/2}
   \Bigg(
   1+ \frac{3}{10(3-\beta)(2-\beta)}\Rcal_h
         \bigg[
              (1-2\beta)\left(\frac{r}{r_h}\right)^{3-\beta}
\nonumber
              \\
           &+  3\left(3-\beta\right)\left(\frac{r}{r_h}\right)  
           -5\left(2-\beta\right)
        \bigg]
    \Bigg).
\end{align}
A similar expression can be easily given for the velocity dispersion.
  
We see from Eq.~(\ref{rhoana}) for plausible stellar density
profiles with $\beta < 3$ that deep inside the cusp
where $r/r_h \ll 1$ the DM density is approximately
  \begin{equation}
 \frac{\rho}{\rho_h} = \left(\frac{r_h}{r}\right)^{3/2}
   \Bigg(
   1 - \frac{3\Rcal_h}{2(3-\beta)}
    \Bigg),\label{rho-limit}
  \end{equation}
i.e. it assumes the same power-law profile $\rho \sim r^{-3/2}$ that we found
in the Fokker-Planck analysis away from the inner boundary 
at $r = r_{cut}$ (see Eq.~(\ref{densfok})). 
By Eq.~(\ref{hydroeq}) we also
get the same velocity dispersion deep inside the cusp,
$v^2 = 3 \vhat^2 = (6/5) M_{bh}/r$ (see Eqs.~(\ref{velfok}) and (\ref{Q})). We also 
find, using Eq.~(\ref{Rcal}), that in the cusp 
the solution asymptotes to
the adiabatic Bondi solution as $r$ decreases, and is adiabatic
{\it everywhere} if ${\Rcal}_h \ll 1$. 

In our Fokker-Planck treatment it was possible to choose the flux in energy space to make $f(E_{cut}) = 0$.  In the two-fluid approach it is tempting to adjust the mass-flux (or, equivalently, $\Rcal_h$) to make $\rho(r_{cut}) = 0$.  While Eq.~(\ref{rho-limit}) shows that $\rho(r)$ at fixed $r$ decreases as $\Rcal_h$ increases, it cannot be shown to make $\rho$ vanish within the perturbation theory, which assumes that $\Rcal_h$ is small.  But even if we were able to solve the present fluid equations without approximation, such a boundary condition would be problematic: requiring $\rho(r_{cut})=0$ with a non-zero mass flux contradicts our assumption that the flow is very subsonic.

To summarise at this point: we note that the two-fluid approach, in contrast to
our Fokker-Planck treatment, does {\it not} yield a
unique value for the DM mass accretion rate, $\dot M$. This fact
is reminiscent of steady-state, adiabatic Bondi flow, for which $\dot M$
is a free parameter that yields viable accretion solutions for all
values up to a maximum, ${\dot M}_{max}$, that depends on $\gamma$. 
We will return to this issue in the next section, once we
have solved the fluid equations without simplifying approximations.  

Before proceeding, however, we make one further observation. 
One way to impose a
reduction in the fluid density at
$r_{cut}$, having imposed boundary conditions at $r_h$, would be to
add a sink term on the right-hand side of Eq.~(\ref{cont}) 
to effectively cut down the DM density inside $r_{cut}$. 
While the DM density and flux are not reduced at $r_{mb}$ when
DM is treated as a fluid, they can be reduced by annihilations. Hence
one could introduce a collision term on the right-hand side
to account for annihilations that would become important 
inside $r \lesssim r_{ann}$, or even a  sink term to model the effect of the loss cone, 
analogous to that introduced in the 
isotropic Fokker-Planck equation in \cite{LigS77,Mer04}.
However, implementing these modifications is beyond the scope of this 
paper, and we shall leave it for a future investigation.

\subsection{Exact Numerical Solution}

The basic fluid equations~(\ref{cont2})-(\ref{energy2}) are
recognized as the usual steady-state, spherical Bondi flow equations with
a heating term on the right-hand side of (\ref{energy2}).
We recently have worked with a similar set of equations, 
but in a different context~\cite{BenGTS19}, namely the accretion of 
baryonic gas accreting onto a SMBH 
(e.g. Sgr A*) heated by DM annihilation. 
Adapting that ``heated Bondi accretion" formalism to the problem at hand, 
we can recast the nonadiabatic fluid equations as follows:
\begin{eqnarray}
\label{bondi1}	
        \frac{d K_{D}}{dr} &=& 	  
	-\frac{(\gamma-1) \Gamma}{\rho^{\gamma} u}, \\
\label{bondi2}
	\frac{d \rho}{d r} &=& -\rho \frac{D_2+H}{D}, \\
\label{bondi3}
	   u &=& \frac{\dot M}{4 \pi \rho r^2},
\end{eqnarray}
where 
\begin{eqnarray}
	D_2 &=& \frac{2 u^2}{r}-\frac{M_{bh}}{r^2}, \\
	D &=& u^2 - a^2, \\
	H &=& \frac{(\gamma - 1) \Gamma}{\rho u},
\end{eqnarray}
and where $a$ is the sound speed,
$P = \rho \vhat^2 = K_{D} \rho^{\gamma}$, $\gamma = 5/3$,
and $\Gamma$ is again given by Eq.~(\ref{energy}). 

In the absence of heating, $\Gamma =0$, $K_D = $ constant, 
and the solution reduces to steady-state, adiabatic Bondi flow
onto a point mass $M_{bh}$ for $\gamma=5/3$. In this case
${\dot M}$ is an eigenvalue which yields valid solutions for 
{\it all} values in the range $0 \leq {\dot M} \leq {\dot M}_{max}$, 
where
\begin{equation}
	\label{dotMmax}
	{\dot M}_{max} = 4 \pi \rho_h u_h r_h^2 =
	   4 \pi \lambda M_{bh}^2 \rho_{\infty} a_{\infty}^{-3},
	   \ \ \ \lambda = 1/4.
\end{equation}
and where the second equality assumes that the fluid is at rest and
homogeneous at infinity. The solution with ${\dot M} = {\dot M}_{max}$
is the only one with $\gamma = 5/3$ that passes through a 
critical transonic point, at which $u=a$. This point is only 
reached at $r=0$, while for all $r>0$ the flow remains subsonic.
For all other ${\dot M} < {\dot M}_{max}$ the flow is  subsonic
everywhere.

As described above, the Newtonian, adiabatic, 
steady-state Bondi equations do {\it not} determine ${\dot M}$ uniquely.
However, the general relativistic analogue 
of these equations for
spherical flow onto a Schwarzshild black hole
shows that the flow {\it must} pass through a critical point to preserve
the causality constraint $a^2 < 1$ and hence this constraint singles out 
flow with $\dot M = {\dot M}_{max}$ as the unique solution for
steady-state flow~\cite{ShaT83}. Furthermore, typical 
{\it time-dependent} integrations 
for adiabatic, spherical accretion 
(i.e. Eqs.~(\ref{cont})-(\ref{energy}) with $\Gamma = 0$)
settle on ${\dot M}={\dot M}_{max}$ when allowed to reach steady-state, 
even in the Newtonian case. 

We have integrated Eqs.~(\ref{bondi1})-(\ref{bondi2}) inward numerically
from $r=r_h$,
adopting the same physical values used in Section IIB in 
our Fokker-Planck treatment for the (outer) boundary conditions 
required by the ODEs for the variables $K_h$ and $\rho_h$ that we 
set at $r = r_h$. We set $\beta = 1.4$ for the background 
stellar density profile, $\rho_* \sim r^{-\beta}$ and $v_h = v_0 = v_{*h}$.

\subsubsection{Flux}

We have considered four cases
for the mass accretion rate $\dot M$, which, as in the case for 
adiabatic Bondi flow, is not determined uniquely in steady-state. 
In particular, we treat
\begin{equation}
	\label{mdotq}
	{\dot M} = q \frac{M_{DM}}{t_{rh}}
\end{equation}
where $t_r$ is  defined in Eq.~(\ref{tr}),
$M_{DM}$ is defined just below Eq.~(\ref{fluxapprox}) and
where we considered four values of $q$ in the range $0.1 \leq q \leq 100$.
The chosen range for $\dot M$ was motivated by the (unique) value expected
from a fundamental Fokker-Planck treatment of the problem 
that solves for $f(E,J)$, as discussed in Section IID 
(see Eq.~(\ref{flux2d})).

Comparison of Eq.~(\ref{dotMmax}) and (\ref{mdotq}) shows that 
\begin{equation}
	\label{fluxratio}
	\frac{\dot M}{\dot M_{max}} \sim q \frac{t_{dynh}}{t_{rh}},
\end{equation}
where $t_{dynh} = r_h/v_h$ is defined as the dynamical
(crossing) timescale at $r_h$. { Evaluating $\dot M_{max}$ here and below we set $a_{\infty} \sim a_h$ and 
$\rho_{\infty} \sim \rho_h$, as in the Bondi solution.} The computed values for 
the adopted Galactic
parameters (Sec.IIB) are $t_{dynh} \sim 2.8 \times 10^3~$yrs, 
$t_{dynh}/t_{rh} \sim 2 \times 10^{-6}$ and 
${\dot M}/{\dot M_{max}} \sim q \times 1.1 \times 10^{-6}$.  
Thus the
anticipated Fokker-Planck accretion rate for which $q \sim 1$ 
is six orders of magnitude smaller than the likely maximum fluid rate.

\subsubsection{Density}

\begin{figure}
\includegraphics[width=7cm]{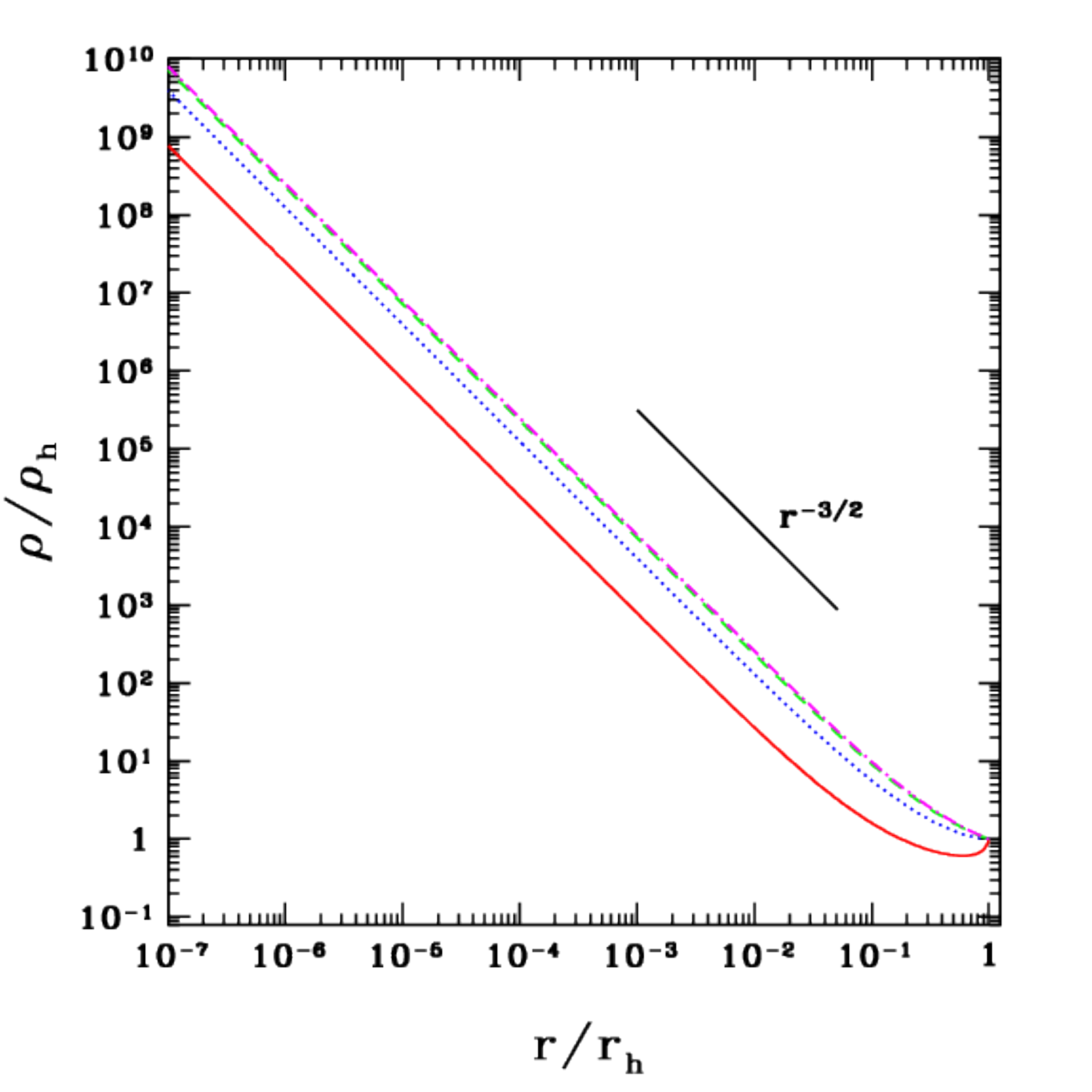}
\caption{Two-component fluid solution for the DM steady-state density
profile $\rho(r)$ in the spike around a massive black hole,
allowing for background stars.  The stellar density is given by 
$\rho_* \sim r^{-\beta}$, with $\beta = 1.4$. Four accretion
rates are chosen according to Eq.~(\ref{mdotq}), with
$q=$0.1 (solid, red); 1 (dotted, blue); 10 (dashed, green); 
100 (dot-dashed, magenta).
The densities and radii are normalized to their values at the
spike outer boundary at $r_h$. Parameters are chosen 
that characterize a spike around Sgr A* in the 
Galactic center (see text).
}
\label{fig:dens}
\end{figure}

\begin{figure}
\includegraphics[width=7cm]{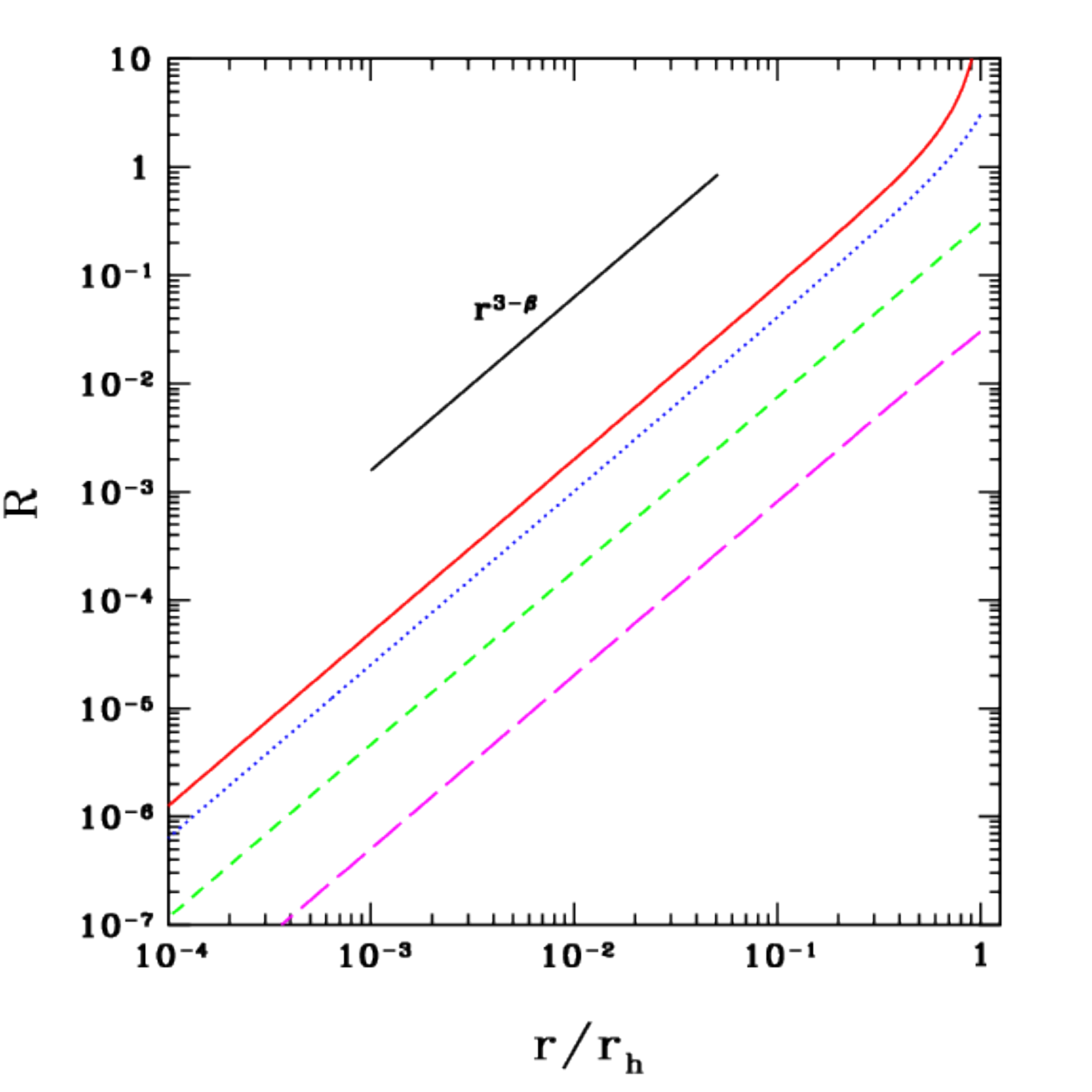}
\caption{Two-component fluid solution for the DM steady-state
dimensionless heating parameter ${\cal R}(r)$ in the spike
around a massive black hole, allowing for background stars
(see Eq.~(\ref{RcalE})).
Results are plotted for the cases
shown in Fig.~\ref{fig:dens} and the labelling is the
same as in that figure.
}
\label{fig:R}
\end{figure}

Results for the DM density profile are plotted in Fig.~\ref{fig:dens}
for all four cases. 
The density satisfies $\rho \sim r^{-3/2}$ for $r \ll r_h$ in all cases.
Moreover, for high values of $q$ and $\dot M$ the profile
obeys this power-law for almost all $r$. This result is not surprising, since
Fig.~\ref{fig:R} shows that the nondimensional heating ratio 
${\cal R} \sim r^{(3-\beta)} \ll 1$ for $r \ll r_h$ in all cases. Since
${\cal R}_h = 3.0/q$,
for sufficiently high $q \gg 1$, and thus high $\dot M$, the ratio 
is small everywhere, even at $r=r_h$. In the latter case the flow is
essentially adiabatic and reduces to the standard adiabatic 
Bondi solution for $\gamma = 5/3$. Our approximate analytic 
profile ~(\ref{rhoana}) reproduces this behavior 
in the perturbative regime.

\subsubsection{Velocity Dispersion}

\begin{figure}
\includegraphics[width=7cm]{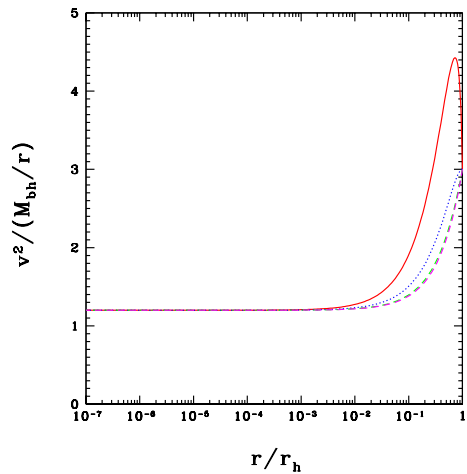}
\caption{Two-component fluid solution for the DM steady-state 
velocity dispersion profile $v(r)$ in the spike 
around a massive black hole, allowing for background stars. 
Results are plotted for the cases shown in 
Fig.~\ref{fig:dens} and the labelling is the
same as in that figure. The velocity dispersion is normalized
to the square of the local circular velocity $M_{bh}/r$.
}
\label{fig:vel2bondi}
\end{figure}

The DM velocity dispersion is plotted in Fig.~\ref{fig:vel2bondi} for
the cases shown in Fig.~\ref{fig:dens}. As expected 
(Sec.IIC), well inside the
outer boundary we find $v^2/(M_{bh}/r) \approx 6/5$. Near $r_h$
the role of heating is reflected in the higher values of
$v^2$ for cases with lower accretion rates. As the accretion rate
is chosen smaller and the corresponding ratio $\cal R$ increases well above
unity, the higher heating rate may subsequently 
unbind the outer regions of the cusp altogether. These 
solutions may then be unstable and a time-dependent integration 
of the equations might then drive the flow to smaller accretion values 
before settling into steady-state.

\subsubsection{Mean Flow Velocity}

\begin{figure}
\includegraphics[width=7cm]{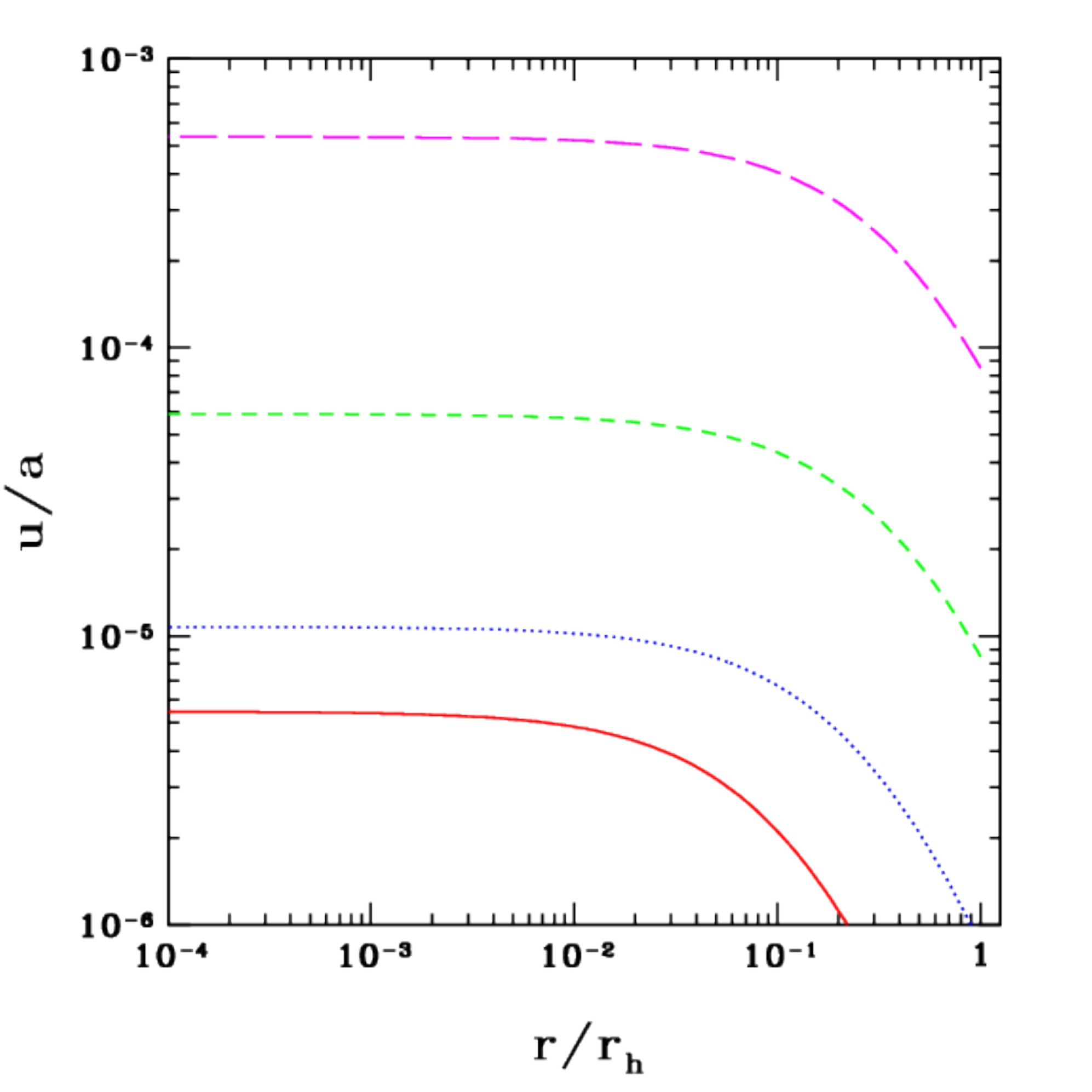}
\caption{Two-component fluid solution for the DM steady-state
Mach number $u(r)/a(r)$ in the spike
around a massive black hole, allowing for background stars.
Results are plotted for the cases
shown in Fig.~\ref{fig:dens} and the labelling is the
same as in that figure. 
}
\label{fig:uova}
\end{figure}

All of our solutions are highly subsonic, as shown in 
Fig.~{\ref{fig:uova}}. This behavior is expected since even 
in adiabatic Bondi flow when $\gamma=5/3$, the mean inflow velocity is
everywhere subsonic, except when ${\dot M} = {\dot M}_{max}$, in
which case $u/a$ reaches unity, but only at the origin. As shown
in Fig.~{\ref{fig:uova}}, the lower the rate of accretion and the
higher the corresponding value of $\Rcal$, the more important heating
becomes and the lower the Mach number $u/a$. Here $a/v = (\gamma/3)^{1/2}
\approx 0.745$ while for $r \ll r_h$ we have $u/a \sim {\rm few} \times 
q (t_{dynh}/t_{rh}) \sim {\rm few} \times 10^{-6}q$, 
which is roughly consistent with Fig.~{\ref{fig:uova}}.

\section{Discussion}

Spherical accretion onto a BH by collisionless matter 
undergoing repeated, small-angle, gravitational scattering is 
qualitatively different from accretion of fluid matter. In the 
former case most of the captured particles
move on highly eccentric orbits that have apocenters far from the 
central hole and are scattered into a loss-cone and captured in one
period. In the latter case, the captured gas moves radially 
as a continuous fluid, becoming tightly bound to the black hole before 
plunging in. All nonradial motion is damped in the case of spherical
fluid flow ~\cite{ShaT83}. Not surprisingly, the accretion rates calculated
by treating DM by these two different descriptions result in two different
answers.

The steady-state rate of accretion anticipated from
a Fokker-Planck treatment of $f(E,J)$,
i.e. Eq.~(\ref{mdotq}) with $q \sim 1$, is orders of magnitude
less than $\dot{M}_{max}$ for adiabatic 
Bondi flow given by Eq.~(\ref{dotMmax}).
The simple flux ratio given by Eq.~(\ref{fluxratio}) highlights this fact.
Yet heating is likely to be unimportant ($\Rcal \ll 1$) well inside
the spike. Hence we anticipate that as the flow approaches the BH, 
a general relativistic treatment  will likely pick out $\dot{M}_{max}$
as the steady-state solution, just as it does for the equations
describing adiabatic Bondi flow, to which the DM fluid equations reduce
deep inside the spike and near the black hole.
Even a time-dependent Newtonian
integration of the equations is likely to relax to this solution.
However, this difference in the predicted accretion rate 
should not lead to a major discrepancy in the
computed DM density or velocity dispersion profiles. We have
already seen that we obtain the same basic power-law 
profiles well inside the spike {when comparing the two-fluid solution 
to the Fokker Planck profiles 
associated with an isotropic $f(E)$.}
Similar agreement is expected when we compare with the
profiles assciated with an anisotropic $f(E,J)$, up to slowly varying logarithmic factors, 
as was proven to be the case for stars in a BW cusp around a BH.

The agreement between Fokker-Planck and fluid profiles breaks down only 
near the outer boundary whenever we have $R_h \gtrsim 1$,
as well as near the inner boundary, since there additional conditions 
can be imposed as inner boundary conditions in the Fokker-Planck solution 
to constrain the distribution function. Constraining the fluid profile
similarly requires the addition of sink terms in the continuity equation,
a departure from the standard two-fluid equations.

Generalizing from this and earlier analyses 
(e.g.~\cite{BetI85,HegA92,SpuT95}), 
of multi-component, large $N$-body dynamical systems undergoing 
secular evolution on relaxation timescales due to gravitational 
scattering, we infer that 
the multi-component fluid approach yields similar results 
to a fundamental Fokker-Planck treatment in many important aspects, 
but not all, depending on the system. One must bear this in mind when
adopting what is often a computationally simpler 
fluid description to describe such a system.

\medskip 

{\it Acknowledgments}: It is a pleasure to thank R. Spurzem for useful
discussions. This paper was supported in part by NSF Grant PHY-2006066
and NASA Grant 80NSSC17K0070 to the University of Illinois at
Urbana-Champaign.

\appendix

\section{Solution of perturbed fluid equations}\label{A1}

The purpose of this appendix is to derive Eq.~(\ref{rhoana}) for the density in the case of weak heating.  The equations to be solved are Eqs.~(\ref{energy2}) and (\ref{hydroeq}), and the right side of the former represents heating.  We regard this term as a perturbation of the no-heating exact solutions of Eq.~(\ref{exact1}), which we denote by $\rho_0$, $\vhat_0$, respectively.  Thus we write the perturbed solution as $\rho = \rho_0(1+f), \vhat = \vhat_0(1+g)$, where $f,g$ are functions of order $\Rcal$.  Substituting into Eq.~(\ref{energy2}), and retaining terms only up to first order in $\Rcal$, we have
\begin{equation}
  -\frac{d}{d\ln r}\left(\ln\left(\frac{\vhat_0^3}{\rho_0}\right) + 3g - f\right) = \frac{3}{2}\Rcal.
\end{equation}
The 0-order term vanishes, as the functions $\rho_0,\vhat_0$ solve the unheated equation exactly, which leads to
\begin{equation}
 -r(3g'-f') = (3/2)\Rcal_h(r/r_h)^{3-\beta},\label{deriv1}
\end{equation}
where we have used Eq.~(\ref{RcalE}), and a prime denotes an $r$-derivative.  This integrates to
\begin{equation}
  3g - f = -\frac{3\Rcal_h}{2(3-\beta)}\left(\frac{r}{r_h}\right)^{3-\beta} + C,\label{integral}
\end{equation}
where $C$ is a constant of integration.

In much the same way, Eq.~(\ref{hydroeq}) gives
\begin{equation}
  f + 2g + \frac{\rho_0\vhat_0^2}{\left(\rho_0\vhat_0^2\right)'}(f'+2g') = f,
\end{equation}
and so, by Eq.~(\ref{exact1}),
\begin{equation}
 -\frac{5}{r}g + f' + 2g' = 0.\label{deriv2}
\end{equation}
Next, Eqs.~(\ref{deriv1}) and (\ref{integral})  let us remove $g$ and $g'$ from Eq.~(\ref{deriv2}), whence
\begin{equation}
  f' - \frac{f}{r} = \frac{3(1-2\beta)}{10(3-\beta)}\frac{\Rcal_h}{r_h}\left(\frac{r}{r_h}\right)^{2-\beta} + \frac{C}{r}. 
\end{equation}

By trying power-law solutions $f\propto r^{\lambda}$ for $\lambda = 3-\beta, 0$ and $1$ 
and superposing, we obtain the general solution
\begin{equation}
  f = \frac{3}{10}\frac{1-2\beta}{(3-\beta)(2-\beta)}\Rcal_h\left(\frac{r}{r_h}\right)^{3-\beta} - C +Dr,
\end{equation}
where $D$ is another constant. The two constants $C$ and $D$ can be chosen so that both $f$ and $g$ vanish
at $r=r_h$, which yields 
\begin{equation}
C = \frac{3}{2(3-\beta)}{\Rcal_h}, \ \ \ 
D = \frac{6(2+\beta)}{10(3-\beta)}\frac{\Rcal_h}{r_h},
\end{equation}
obtaining Eq.~(\ref{rhoana}).  

\bibliography{paper}
\end{document}